\documentclass[prl,twocolumn,preprintnumbers,superscriptaddress,amsmath,amssymb, longbibliography]{revtex4-1}
\usepackage{graphicx}
\usepackage{mathrsfs}
\usepackage{amsfonts}
\usepackage{times}
\usepackage{amsmath}
\usepackage{leftidx}
\usepackage{tikz}

\usepackage[colorlinks,linkcolor=blue,citecolor=blue]{hyperref}
\usepackage{bbold}
\usepackage{braket}
\usepackage{mathtools}

\newcommand{\kvec}{{\bf k}}
\newcommand{\sset}{{\bm \sigma}}

\newcommand{\bigO}{\mathcal{O}}

\begin{document}
\title{Solving Quasiparticle Band Spectra of Real Solids using Neural-Network Quantum States}
\author{Nobuyuki Yoshioka}
\email[E-mail at: ]{nyoshioka@ap.t.u-tokyo.ac.jp}
\affiliation{Theoretical Quantum Physics Laboratory, RIKEN Cluster for Pioneering Research (CPR), Wako-shi, Saitama 351-0198, Japan}
\affiliation{Department of Applied Physics, University of Tokyo, 7-3-1 Hongo, Bunkyo-ku, Tokyo 113-8656, Japan}
\author{Wataru Mizukami}
\affiliation{Center for Quantum Information and Quantum Biology, Institute for Open and Transdisciplinary Research Initiatives,
Osaka University, Osaka 560-8531, Japan}
\affiliation{JST, PRESTO, Kawaguchi, Saitama 332-0012, Japan}
\affiliation{Graduate School of Engineering Science, Osaka University, Toyonaka, Osaka 560-8531, Japan}
\author{Franco Nori}
\affiliation{Theoretical Quantum Physics Laboratory, RIKEN Cluster for Pioneering Research (CPR), Wako-shi, Saitama 351-0198, Japan}
\affiliation{Physics Department, University of Michigan, Ann Arbor, Michigan 48109-1040, USA}
\date{\today}

\begin{abstract}
{\it Abstract.---} Establishing a predictive ab initio method for solid systems is one of the fundamental goals in condensed matter physics and computational materials science. 
The central challenge is how to encode a highly-complex quantum-many-body wave function compactly. Here, we demonstrate that artificial neural networks, known for their overwhelming expressibility in the context of machine learning, are excellent tool for first-principles calculations of extended periodic materials. 
We show that the ground-state energies in real solids in one-, two-, and three-dimensional systems are simulated precisely, reaching their chemical accuracy. 
The highlight of our work is that the quasiparticle band spectra, which are both essential and peculiar to solid-state systems, can be efficiently extracted with a computational technique designed to exploit the low-lying energy structure from neural networks.
This work opens up a path to elucidate the intriguing and complex many-body phenomena in solid-state systems.

\end{abstract}
\maketitle

\section{Introduction}
Artificial neural networks (ANNs) are a class of expressive mathematical models originally designed to imitate the high computing power of the human brain. 
Driven by the outstanding success over existing data processing methods in the field of machine intelligence~\cite{krizhevsky_2012, goodfellow_2016, silver_2016}, 
ANNs have been used in a wide range of applications, from physical science~\cite{carleo_2019, sarma_2019, carrasquilla_2020, carrasquilla_2017, Yoshioka_Learning_2018}, medical diagnosis, to astronomical observations.
Remarkable among numerous factors underlying their performance is their ability to perform efficient feature extraction from high-dimensional data.

\begin{figure}[t]
    \begin{center}
    \hspace{-1cm}
     \resizebox{0.85\hsize}{!}{\includegraphics{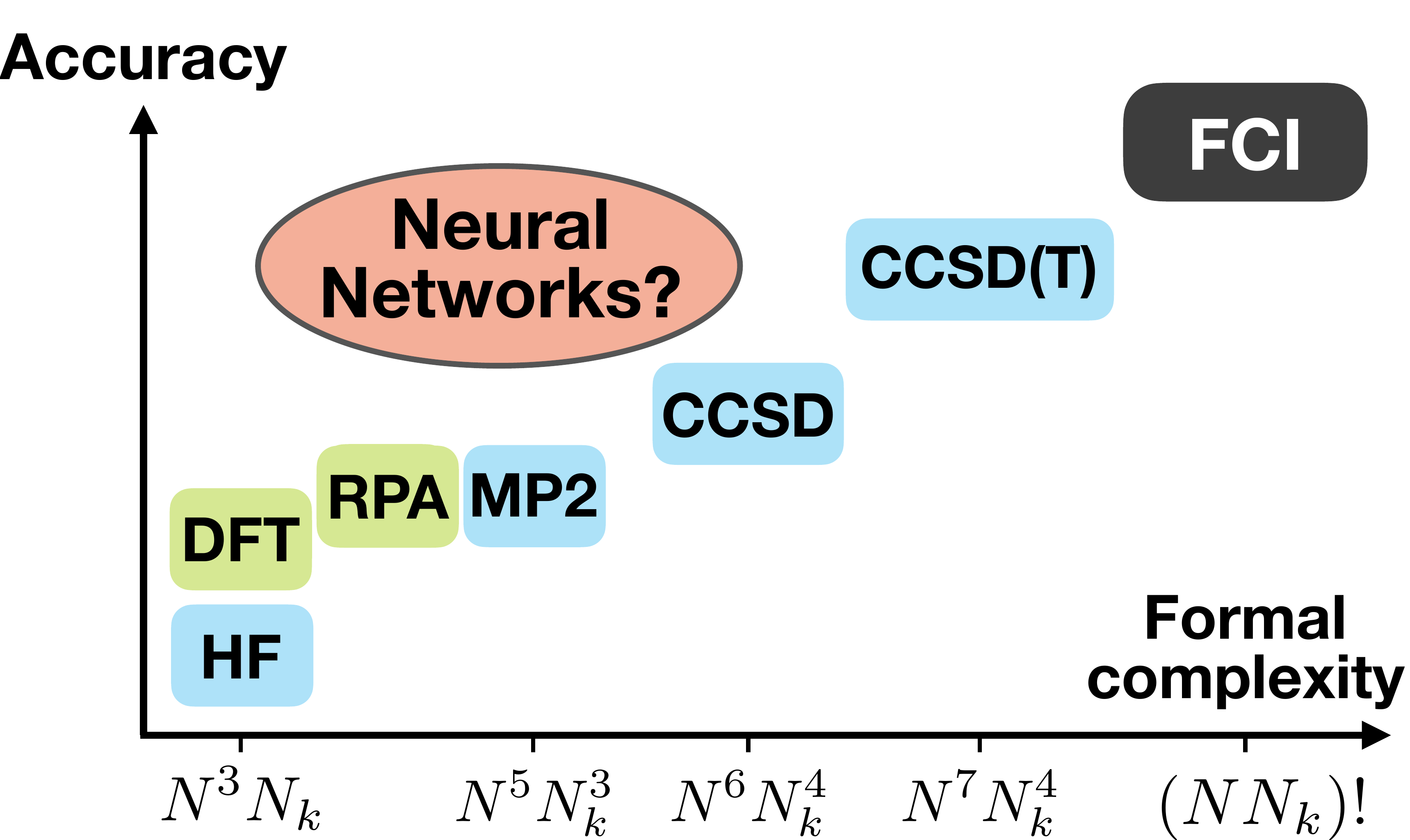}}
   \end{center}
\caption{\label{fig:fig1} {\bf Schematic illustration of the relationship between the formal computational complexity and accuracy in various first-principles calculation methods for solid systems.} 
Our goal is to demonstrate that the variational calculation using neural-network-based ansatz can readily describe both weakly and strongly correlated electronic structures with moderate number of variational parameters, i.e., computational cost.
We denote the full configuration interaction (FCI) method by the black square, whereas the Hartree--Fock (HF) and post-HF calculation methods are indicated by blue squares:
the second-order M\o ller--Plesset perturbation theory
(MP2), the coupled-cluster singles and doubles (CCSD), and CCSD with perturbative triple excitations (CCSD(T)).
Also, the green squares indicate methods based on the Density Functional Theory (DFT): the DFT and DFT-based Random Phase Approximation (RPA). 
The number of orbitals at each $k$-point is denoted as $N$ and the total number of $k$-points as $N_k$. Note that this is a qualitative (approximate) illustration which will vary from case to case.
}
\end{figure}


As universal approximators, ANNs have a rich expressive power, which can also be exemplified by encoding complicated quantum correlations~\cite{schuld_2014}.
Reference~\cite{carleo_2017} showed that ANNs, employed as a quantum many-body wavefunction ansatz, can solve strongly correlated lattice systems at state-of-the-art level.
Such quantum state ansatze, often referred to as neural quantum states (NQS), capture quantum entanglement that even scales extensively~\cite{deng_prx_2017}.
The use of such a powerful non-linear parametrization has been keenly investigated in the quantum physics community: both equilibrium~\cite{nomura_2017, deng_prb_2017} and out-of-equilibrium~\cite{yoshioka_2019, hartmann_2019, nagy_2019, vicentini_2019} properties,  extension of the network structure~\cite{gao_2017,  choo_2018, levine_2019}, and quantum tomography~\cite{torlai_tomography_2018, torlai_NDO_2018, melkani_2020, ahmed_2020}.
Meanwhile, we point out that the application of ANNs to fermionic systems is much less explored,
despite their practical significance, such as the modeling of real materials and the experimental realizability in quantum simulators~\cite{georgescu_2014, mazurenko_2017}. 
\textcolor{black}{The proof of concept for small molecular systems was first presented by Choo {\it et al.}~\cite{choo_2020} which applied the ANNs to solve the many-body Schr\"{o}dinger equation governed by the second-quantized Hamiltonian for molecular orbits. 
Few implementations have been further performed to simulate the electronic structures using ANNs~\cite{yang_2020, pfau_2020, hermann_2020, spencer_2020}.
Thus, a crucial question remains to be answered: are ANNs powerful enough to represent the electronic structures of real solid materials?} 
This is related to one of the fundamental problems in condensed matter physics and computational materials science; namely, establishing a predictive ab initio method for solids or surfaces. 
\textcolor{black}{In particular, it must be demonstrated that the ANNs are capable of investigating the thermodynamic limit.
}

We stress that no current first-principles method can take into account both weak and strong electron correlations compactly and sufficiently.  
For instance, it is well known that the accuracy of the de facto standard method, density functional theory (DFT), is semi-quantitative and it is very difficult to improve significantly~\cite{Medvedev2017Science,Mardirossian2017MolPhys}.
Many-body-wave-function-based methodologies are, in contrast, systematically improvable.
Such techniques, mainly based on coupled cluster (CC) theory (or many-body perturbation theory)~\cite{Shavitt2009}, have been successful for the electronic states of molecules.
This has encouraged the application of quantum chemical methods to solid state physics~\cite{Gruber2018PRX,Zhang2019Frontiers}.
However, methods such as CC specialize in describing weak electronic correlations, and only work well for electronic states where the mean-field approximation is valid.

Methods for dealing with strongly correlated electrons, called multi-reference theory, also exists in quantum chemistry~\cite{Roos2016Book}; but these assume that the number of strongly correlated electrons is small.
Such a condition usually holds in the case of molecules, because the number of strongly correlated electrons is often localized and limited.
In contrast, there can be a large number of moderately or strongly correlated electrons in solid-state systems, owing to their high symmetry and dense structure.
Based on its success in spin systems, it is natural to expect that the NQS have the potential to compactly describe a variety of electron correlations appearing in first-principles calculations of solids with a moderate computational cost
(See Fig.~\ref{fig:fig1} for a schematic diagram of the hierarchy of quantum chemical methods\textcolor{black}{~\cite{hirata_2004, zgid_2011, ke_2016, mcclain_2017, booth_2013}}).

\textcolor{black}{In this work, we demonstrate that neural-network-based many-body wave functions can readily simulate the essense of first-principles calculations for extended periodic materials: the ground-state and excited-state properties.}
The second-quantized fermionic Hamiltonian is transformed into a spin representation, such that the problematic sign structure of fermions, which usually imposes severe limits on the numerical accuracy, is naturally encoded.
Employing the variational Monte Carlo (VMC)-based stochastic optimization, we show that the thermodynamic limit of a one-dimensional system
can be simulated within chemical accuracy.
For real solids in both two and three dimensions, the static electronic correlation in the minimal active space is compactly represented by the NQS.
\textcolor{black}{Our work's main contribution is that multiple excited states, forming quasiparticle band spectra, are computed by constructing an effective Hamiltonian in the truncated Hilbert space. 
To the best of our knowledge we offer the first demonstration that the NQS can be applied to simulate low-lying eigenstates in the identical-quantum-number sector.
}

\section{Results}
\subsection{Second quantization representation of solid systems}
To alleviate the notorious difficulty of simulating the many-body problem of solid systems, we employ a linear combination of the single-particle basis. 
Namely, we construct crystalline orbitals (COs) using the solution of the crystalline Hartree--Fock (HF) equation~\cite{del_1967, andre_1969}. 
The second-quantization form of the many-body fermionic Hamiltonian is
\begin{eqnarray}\label{eqn:fermionic_hamiltonian}
    H &=& \sum_{pq} \sum_{\kvec} t_{pq}^{\kvec}c_{p\kvec}^{\dagger}c_{q\kvec} \nonumber \\
    &+& \frac{1}{2}\sum_{pqrs}\sum_{\kvec_p \kvec_q \kvec_r \kvec_s}' v_{pqrs}^{\kvec_p \kvec_q \kvec_r \kvec_s} c_{p\kvec_p}^{\dagger} c_{q\kvec_q} c_{r \kvec_r}^{\dagger} c_{s\kvec_s},
\end{eqnarray}
where $c_{p\kvec}$ ($c_{p\kvec}^{\dagger}$) denotes the annihilation (creation) operator of an electron on the $p$-th CO with crystal momentum $\kvec$. 
Here, the anticommutation relation $\{c_{p \kvec_p}, c_{q\kvec_q}^{\dagger}\} = \delta_{pq}\delta_{\kvec_p \kvec_q}$ is imposed, and
one-body (two-body) integrals are given as  $t_{pq}^{\kvec}$ ($v_{pqrs}^{\kvec_p\kvec_q\kvec_r\kvec_s}$). 
For simplicity, hereafter we denote the suffix as $\mu:=(p\kvec)$.
While the general framework of the crystalline HF equation is common with that for molecular systems, it must be noted that the contribution from the reciprocal lattice vector ${\bf G} = 0$ requires extra numerical care owing to the divergence of the exchange integrals.
In this work, we employ the crystalline Gaussian-based atomic functions as the single-particle basis. 
The Gaussian density fitting technique is applied to efficiently compute the two-body integrals~\cite{Sun2017JCP_GDF}.

The summation in the first term of Eq.~\eqref{eqn:fermionic_hamiltonian} is taken over a uniform grid, which is typically obtained by shifting the $k$'s obeying the Monkhorst--Pack rule~\cite{monkhorst_1976}.
Note that the number $N_k$ of sampled $k$-points can be arbitrary.
The primed summation in the second term satisfies the conservation of crystal momentum, which follows from translational invariance: 
\begin{eqnarray}
\kvec_p + \kvec_r - \kvec_q - \kvec_s \in \mathcal{G},
\end{eqnarray}
where $\mathcal{G}$ is the set of reciprocal lattice vectors. 
With the number of COs at each $k$-point denoted as $N$, the total number of terms in Eq.~\eqref{eqn:fermionic_hamiltonian} is given as $\bigO(N^4 N_k^3)$.

\begin{figure*}[t]
\begin{center}
\includegraphics[width=1.95\columnwidth]{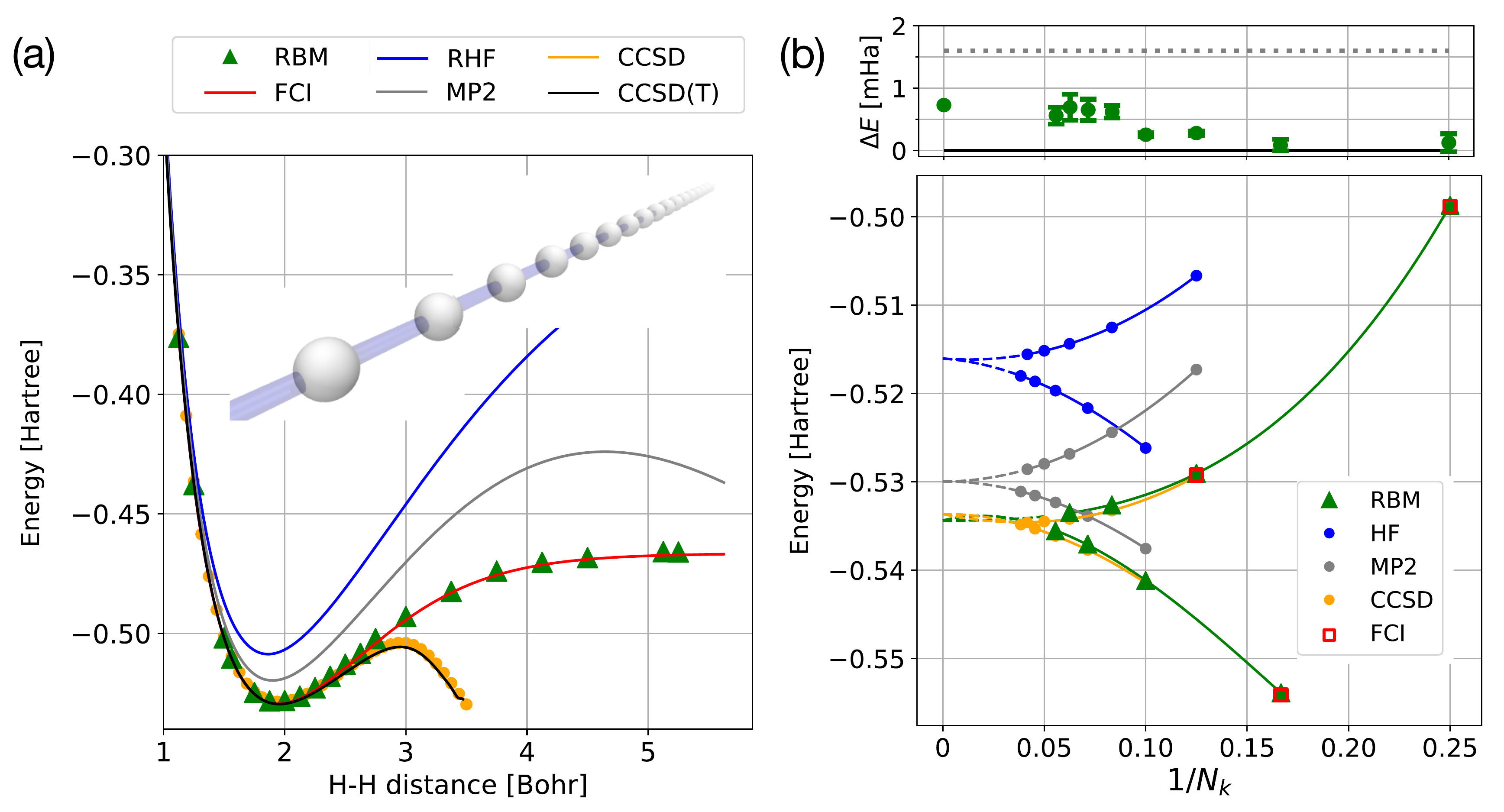}   
\caption{{\bf Solving the ground state of the linear hydrogen chain using the minimal STO-3G basis set.}
(a) The potential energy curve calculated by the restricted Boltzmann machine (RBM) agrees with the full configuration interaction (FCI) method within chemical accuracy (1.6 mHa) for any atom separation $d_H$. 
This indicates that the RBM states are capable of describing both the weakly and strongly interacting regimes, where gold-standard techniques, such as coupled-cluster singles and doubles (CCSD) shown by the yellow line and CCSD with perturbative triple excitations (CCSD(T)) in black line, break down.
The results by restricted Hartree--Fock (RHF) and second-order M\o ller--Plesset perturbation theory are indicated by blue and gray lines, respectively.
A unit cell consists of four hydrogen atoms placed at even intervals, and two $k$-points are sampled from a uniform grid.
(b) Finite-size scaling of the ground state energy up to $N_k = 18$ and its deviation from the FCI, $\Delta E$, at near-equilibrium $d_H = 2$. The results show excellent agreement with conventional methods even in the thermodynamic limit $N_k \rightarrow \infty$. 
Here, the unit cell consists of a single hydrogen atom, and hence the maximum number of spin-orbitals considered here is 36.
The error bars denote the standard deviation of the estimation by the Monte Carlo sampling.
}
\label{fig:hydrogen_chain}
\end{center}
\end{figure*}

To solve the fermionic many-body Hamiltonian \eqref{eqn:fermionic_hamiltonian},
we must explicitly impose the antisymmetric sign structure in the quantum state.
Here, we map the Hamiltonian into the spin-1/2 representation such that the sign structure is encoded in the operators rather than the quantum states, 
\textcolor{black}{as Choo {\it et al.}~\cite{choo_2020} considered in their application of the NQS to small molecules.}
The Jordan--Wigner (JW)  transformation~\cite{jordan_1928} defines the relation of fermionic and spin operators as $c_{\mu}^{(\dagger)} = (-1)^{\mu-1}\prod_{\nu<\mu}\sigma_{\nu}^z \sigma_{\mu}^{+(-)}$,
where $\sigma_{\mu}^{+(-)}$ is the raising (lowering) operator of the $\mu$-th spin.
Such a mapping yields a non-local spin Hamiltonian 
\begin{eqnarray}\label{eqn:pauli_hamiltonian}
    H = \sum_{Q} c_Q P_Q,
\end{eqnarray}
where $P_Q \in \bigotimes_{\mu}\{I, X, Y, Z\}$ is a product of Pauli matrices for a corresponding Pauli string $Q$.

\textcolor{black}{Let us make two remarks on the application of JW transformation.
First, the use of the fermion-to-spin transformation for stochastic variational calculations was initially considered in the context of near-term quantum computers~\cite{peruzzo_2014},
including the application to real solids~\cite{liu_2020, manrique_2020, yoshioka_2020}, while the spin-to-fermion mapping has been long applied in condensed matter and statistical physics community, e.g., to solve exactly soluble quantum spin models.}
Second, the JW transformation merely generates the spin operator representation of the Hamiltonian~\eqref{eqn:fermionic_hamiltonian} and does not alter the computational basis.
The evaluation of physical observables in the Monte Carlo approach by the occupation-number basis of the fermionic representation is identical to that by the spin computational basis of the spin representation.
This is not the case when we apply other transformations developed in quantum information, such as the Bravyi--Kitaev transformation~\cite{bravyi_2002}.

\subsection{Ground states in the thermodynamic limit}
In general, it is classically intractable to solve for the ground state of the many-body Hamiltonian defined in Eq.~\eqref{eqn:fermionic_hamiltonian} or~\eqref{eqn:pauli_hamiltonian}.
Here we alternatively rely on a variational method that exemplifies the expressive power of neural networks.
\textcolor{black}{Namely, a neural network is used as a variational many-body wave function ansatz. 
It is optimized so that the expectation value of the energy, estimated via the Monte Carlo simulation, is minimized by approximating the imaginary-time evolution.
Such a technique, called variation Monte Carlo (VMC), has been successfully applied to condensed-matter systems~\cite{mcmillan_1965, koloren_2011, sorella_2002, misawa_2014} and quantum chemistry problems~\cite{hammond_1994, foulkes_2001}, leading to state-of-the-art numerical analysis on strongly correlated phenomena.
The choice of the  variational ansatz plays a key role for the accuracy, which, as has been pointed out by Carleo and Troyer~\cite{carleo_2017}, can be significantly improved by using neural networks.}

\textcolor{black}{Let us briefly review the general protocol of VMC for simulating ground-states in many-body spin systems using the quantum-state ansatz based on the restricted Boltzmann machine (RBM)~\cite{smolensky_1986}. }
First, we introduce the quantum many-body wave function expressed as follows~\cite{carleo_2017},
\begin{eqnarray}\label{eqn:rbm_state}
    \ket{\Psi^{\rm RBM}_{\theta}} &=& \frac{1}{Z}\sum_{\sset}\Psi^{\rm RBM}_{\theta}(\sset) \ket{\sset}, \nonumber \\
    \Psi_{\theta}^{\rm RBM} (\sset) &=& \sum_{h} \exp(W_{\mu\nu}\sigma_{\mu} h_{\nu} + \sum_{\mu}a_{\mu}\sigma_{\mu} + \sum_{\nu} b_{\nu} h_{\nu}),\nonumber \\
    \ 
\end{eqnarray}
where $\Psi^{\rm RBM}_{\theta}(\sigma)$ is the unnormalized amplitude for a spin configuration $\sigma \in \{-1, +1\}^{N_v}$ where $N_v = N N_k$ is the total number of spin orbitals and $Z=\sqrt{\sum_{\sigma} |\Psi^{\rm RBM}_{\theta} (\sigma)|^2}$ is the normalization factor. 
We denote the set of complex variational parameters as $\theta=\{W_{\mu\nu}, a_{\mu}, b_{\nu}\}$,
where the interaction $W_{\mu\nu}$ denotes the virtual coupling between the spin $\sigma_{\mu}$ and the auxilliary degrees of freedom, or the hidden spin $h_{\nu}$.
One-body terms $a_{\mu}$ and $b_{\nu}$ are also introduced to enhance the expressive power of the RBM state.
In the present work, we find that the it suffices to take the total number of the hidden spin as $N_h = N_v$, and therefore the number of the complex variational parameters is $(N_v^2 + 2N_v)$ in total.
\textcolor{black}{The all-to-all connectivity between $\sigma$ and $h$ allows the RBM state to capture complicated quantum correlations such as topological orders~\cite{deng_prb_2017, glasser_prx_2018}, spin-liquid behaviours~\cite{choo_2019, ferrari_2019, nomura_2020}, and electronic structures in small molecular systems~\cite{choo_2020, yang_2020}.}

Using the RBM state~\eqref{eqn:rbm_state} as the many-body variational ansatz, the ground state problem is solved \textcolor{black}{in the variational Monte Carlo framework.
In particular, we rely on the stochastic reconfiguration technique~\cite{sorella_2001} to approximate the imaginary-time evolution as 
\begin{eqnarray}
    \ket{\Psi_{GS}} \propto \lim_{\tau \rightarrow \infty} e^{-\tau H} \ket{\Psi_0} \sim \ket{\Psi^{\rm RBM}_{\theta_0 + \sum_k \Delta \theta_k}},
\end{eqnarray}
where the parameter update at the $k$-th step $\Delta \theta_k$ is given by the Monte Carlo simulation, and the initial state $\ket{\Psi_0}$ is taken as the HF state in our simulation.}
Detailed information on the implementation and optimization techniques is provided in ``Methods".

\begin{figure*}[t]
\begin{center}
\includegraphics[width=1.95\columnwidth]{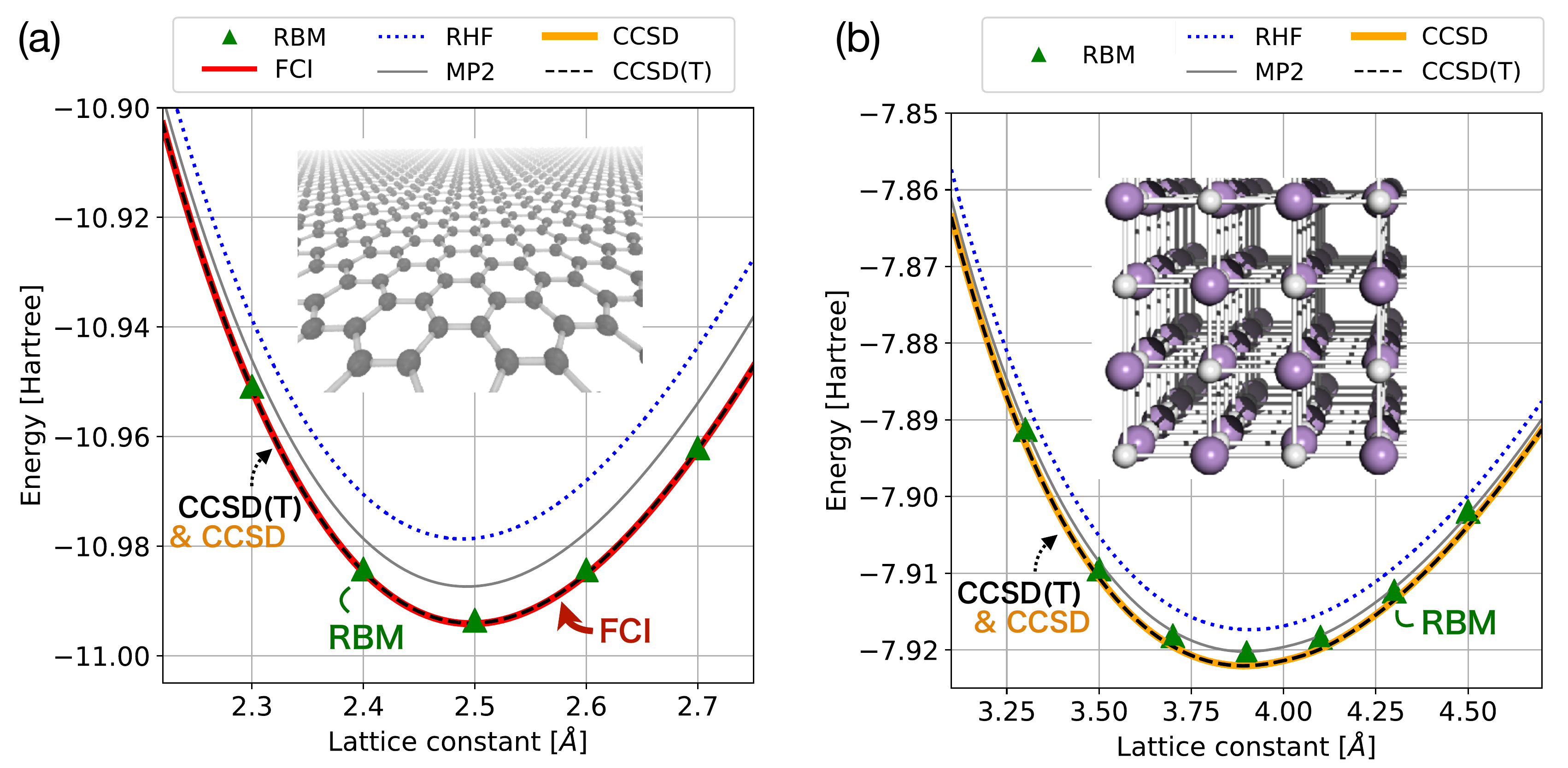}
\caption{{\bf Potential energy curves for 2D and 3D real solids calculated by neural networks. }
The ground state energy is computed for various lattice constants in the vicinity of equilibrium values. (a) Graphene on a honeycomb lattice solved using the cc-pVDZ basis set. The smallest active space is taken at each $2 \times 2$ $\Gamma$-centered $k$-point, and hence 
16 spin orbitals in total. (b) LiH with the rocksalt structure solved using the STO-3G basis.
The smallest active space is taken at each $2 \times 2 \times 2$ $\Gamma$-centered $k$-point, and hence 32 spin orbitals in total.  The result obtained for the RBM state (green triangle) shows remarkable agreement either with the full configuration interaction (FCI) method, achieving an error within chemical accuracy (1.6 mHa). 
The red, blue dotted, gray, yellow, and black dashed lines denote the results by the FCI method, restricted Hartree-Fock (RHF) method, second-order M\o ller--Plesset perturbation theory (MP2), coupled-cluster singles and doubles (CCSD), and CCSD with perturbative triple excitations (CCSD(T)), respectively.
}
\label{fig:graphene_LiH}
\end{center}
\end{figure*}

As a first demonstration, we provide the potential energy curve for a one-dimensional system whose electronic correlation varies drastically as the geometry is changed. 
Concretely,  we consider a linear hydrogen chain with homogeneous atom separation $d_H$ in a minimal basis set (STO-3G)~\cite{motta_2017, motta_2020}.
Figure~\ref{fig:hydrogen_chain}(a) presents the result of the calculation using the RBM state
as well as the second-order M\o ller--Plesset perturbation theory
(MP2)~\cite{sun_1996}, the coupled-cluster singles and doubles (CCSD)~\cite{hirata_2001, mcclain_2017}, and  CCSD with perturbative triple excitations (CCSD(T))~\cite{gruneis_201l}, which is considered as the gold-standard in modern quantum chemistry.
While the weakly correlated regime at near-equilibrium is simulated quite well by all the conventional methods, we see that they start to collapse as the correlation grows at the intermediate $d_H$ regime, not to mention the Mott-insulating large $d_H$ regime.
In sharp contrast, the RBM state precisely  describes the electronic correlation and achieves chemical accuracy at any atom separation $d_H$.
Here, two $k$-points are sampled from each unit cell which contains four hydrogen atoms so that the interactions between nearby sites are reflected explicitly on the model. 

To further illustrate the RBM state's power and reliability, we calculate the energy in the thermodynamic limit by extrapolating $N_k \rightarrow \infty$ in a system with a single atom per unit cell.
The numerical result at near-equilibrium ($d_{H}=2.0 a_B$) is shown in Fig.~\ref{fig:hydrogen_chain}(b).
We confirm the excellent agreement with conventional methods by comparing the result with the FCI for $N_k \leq8$ and CCSD for $10 \leq N_k \leq 18$.
Clearly, the thermodynamic limit is simulated precisely as well as the finite-size system.

Next, we provide the demonstration in both 2D and 3D real solids: graphene and the lithium hydride (LiH) crystal in the rocksalt structure. Here, we restrict the active space per each $k$-point to its highest occupied CO and lowest unoccupied CO.
The results for graphene [Fig.~\ref{fig:graphene_LiH}(a)] and the crystalline LiH [Fig.~\ref{fig:graphene_LiH}(b)] are both in remarkable agreement with the FCI or CCSD(T). 
Clearly, the RBM ansatz gives a quantitatively accurate description which may allow crystal structure determinations of weakly to moderately correlated real solid systems.

\begin{figure*}[t]
\begin{center}
    \includegraphics[width=1.95\columnwidth]{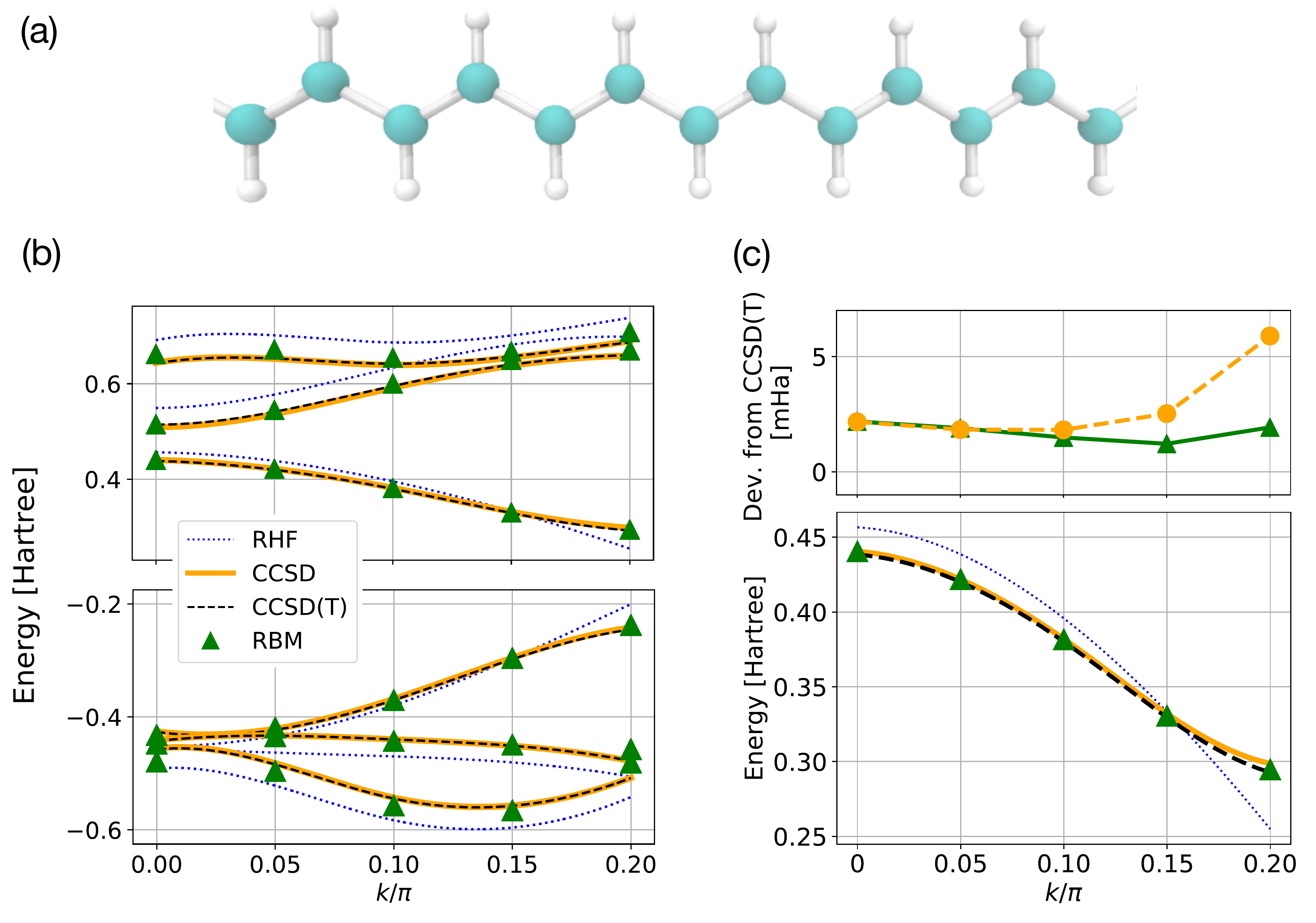}
\caption{\label{fig:C2H2_band} 
{\bf Quasiparticle band spectra from multiple excited-state calculation.}
(a) Schematic diagram of the trans-polyacetylene (C$_2$H$_2$)$_n$. The cyan and gray spheres indicate the carbon and hydrogen atoms, respectively.
\textcolor{black}{
(b) Three quasiparticle bands below and above the Fermi energy. 
Here, the yellow lines and black dashed lines indicate results obtained from the equation-of-motion  coupled cluster (EOM-CC) formalism; CCSD and CCSD(T) stand for the unperturbed EOM-CCSD and the perturbed EOM-CCSD(T)(a)*methods, respectively. The blue dotted lines denote the restricted Hartree--Fock (RHF) method.
(c) A zoom-in of the first conduction band, which is computed from the electron attatchment (EA) energy. 
It is clearly shown, from the energy differences against the EA-EOM-CCSD(T)(a)* method, that the results by the RBM (green triangle) are comparable or better than the unperturbed EA-EOM-CCSD method.
In all calculations, a single $k$-point is taken under the minimal basis set (STO-3G) and hence 24 spin orbitals are taken into account.
The size of the unit cell is taken as 2.451 \AA.
}
}
\end{center}
\end{figure*}


\subsection{Quasiparticle band structure from the one-particle excitation}
Interest beyond the ground-state electronic structures in solids is diverse: the response against electromagnetic fields, impurity effects, phononic dispersions, and so on. 
Here, we focus on the band structure, which is a peculiar yet fundamental property that characterizes solid systems.
We stress that variational calculations for the lowest band gap, which can be experimentally measured from photoemissions, are already few, 
\textcolor{black}{not to mention the simulation of the band spectra based on stochastic methods~\cite{ma_2013}.} 
Furthermore, to the best of our knowledge, there is no NQS simulation of excited states  in the identical sector of quantum numbers except the first excited state~\cite{choo_2018}.
\textcolor{black}{This motivates us to perform the first attempt to calculate multiple low-lying states and deepen our understanding on the representability of the NQS beyond the well-studied regimes.
}

In general, the calculation of band structures is based on the assumption that the system is weakly to moderately correlated.
In other words, the mean-field approximation is qualitatively valid, so that one-particle excitations dominate the low-lying spectrum. 
By employing such a picture in a quantum many-body context, we can also simulate the band structure via quasiparticle excitations. 
We take a similar approach here and compute the band structure from the single-particle linear-response behavior of the ground state.

Let us construct an appropriately truncated Hilbert space which captures the low-lying states in a stochastic manner.
It is justified from the above argument that we consider a subspace spanned by a set of non-orthonormal bases $\{ R_{\alpha}\ket{\Psi_{\rm GS}} \}$, where $R_{\alpha}$ denotes the $\alpha$-th single-particle excitation operator.
Here, the valence (conduction) bands are obtained from the ionization (electron attachment) operators $\{c_{p\kvec_p}\}$ ($\{c^{\dagger}_{p\kvec_p}\}$), which allows us to compute the quasiparticle band with an additional computational cost of $\bigO(N_v^3)$.
\textcolor{black}{Though it is possible to include higher order excitation operators, here we avoid them from the viewpoint of computational cost and size inconsitency.
}
It can be shown that the diagonalization of the effective Hamiltonian given the non-orthonormal basis is done by the following generalized eigenvalue equation~\cite{mcclean_2017},
\begin{equation}
    \widetilde{H} C = SCE,\label{eqn:qse}
\end{equation}
where $E = {\rm diag}(E_1, ..., E_{N_v})$ denote the eigenvalues and $C$ is an array of eigenvectors. 
The matrix elements of the non-hermitian matrix $\widetilde{H}$ and the metric $S$ are estimated via the Monte Carlo sampling as expectation values:
\begin{eqnarray}
    \widetilde{H}_{\alpha\beta} &=& \braket{\Psi^{\rm RBM}_{\rm \theta^*}| R^{\dagger}_{\alpha} H R_{\beta} | \Psi^{\rm RBM}_{\rm \theta^*}}, \label{eqn:Hsub}\\
    S_{\alpha \beta} &=& \braket{\Psi^{\rm RBM}_{\rm \theta^*} | R^{\dagger}_{\alpha} R_{\beta}|\Psi^{\rm RBM}_{\rm \theta^*}},\label{eqn:Ssub}
\end{eqnarray}
where the ground state is now replaced by the RBM ansatz $\ket{\Psi_{\theta^*}^{\rm RBM}}$, with the optimized variational parameter $\theta^*$.
\textcolor{black}{In the field of quantum chemistry, this procedure is referred to as the internally-contracted multireference configuration interaction ~\cite{werner1982self,werner1988efficient}.
}

To enhance the numerical reliability, \textcolor{black}{we incorporate the effect of orbital relaxation by estimating the band gap} from the extended Koopmans' theorem~\cite{day_1974, smith_1975, morrell_1975}. 
The energies are shifted so that the first valence and conduction bands coincide with the energy difference
$\Delta E^{IP}$ and $\Delta E^{EA}$ as
\begin{eqnarray}\label{eqn:koopmans}
\begin{cases}
\Delta E^{IP} &= E_{GS}^{N_v} - E_{GS}^{N_v-1}, \\
\Delta E^{EA} &= E_{GS}^{N_v+1} - E_{GS}^{N_v},
\end{cases}
\end{eqnarray}
where $E_{GS}^{n}$ is the energy of the RBM optimized in the particle-number sector $n$ (See ``Methods").

We provide a demonstration for the quasiparticle band structure of the polyacetylene [Fig.~\ref{fig:C2H2_band}(a)] using the STO-3G basis sets. 
The result is compared with a variant of the equation-of-motion coupled cluster theories (EOM-CC): ionization-potential (electron-attached) EOM-CC (IP-EOM-CC, EA-EOM-CC) which considers up to 2-hole and 1-particle (2-particle and 1-hole) excitations~\cite{mcclain_2017}.
The agreement with EOM-CCSD(T)(a)*~\cite{matthews_2016} is very good for the first valence and conduction bands, while it becomes slightly worse for higher excitations.
As is shown in Fig.~\ref{fig:C2H2_band}(b), the first conduction band is simulated almost within chemical accuracy, which is partly due to the cancellation of the optimization errors induced by Eq.~\eqref{eqn:koopmans}.
Meanwhile, Fig.~\ref{fig:C2H2_band}(c) indicates that errors in the higher excitations can be an order of magnitude larger in the worst case, which cannot be explained merely from the variational simulation error. 
Rather, it can be understood as a systematic error originating in the insufficiency of the truncated Hilbert space;
there is a trade-off between the computational cost and the accuracy.
Systematic improvement can be expected from using higher-order excitation operators, e.g., two-electron excitation operators
$\{c^{\dagger}_{p \kvec_p}c_{q \kvec_q}\}$ for the lowest energy state in the particle-number sectors $(N_v \pm 1)$.

\section{Conclusion}
We have shown that a shallow neural network with a moderate number of variational parameters allows us to perform the essence of first-principles calculations in solid systems, i.e., the ground-state property and the quasiparticle band spectra. 
In the weakly to moderately correlated regions of the linear hydrogen chain, we have demonstrated that even the thermodynamic limit can be simulated using the RBM state.
The representability of the RBM is also exhibited in the strongly correlated regions, where the standard approaches break down.
We have furthermore shown that the electronic structures of real solids in both 2D and 3D can be described accurately.
Furthermore, we have successfully obtained the quasiparticle band spectra of a polymer in the linear response regime.
To the best of our knowledge, this is the first demonstration proving that NQS are capable of computing multiple excited states, in addition to precise ground-state simulations that reach their chemical accuracy.

Numerous future directions can be envisioned. 
We remark the following three points.
First is the extension towards the complete basis limit. 
While we have here focused on relatively simple basis sets, the quantitative prediction and comparison with experiments would necessarily require larger basis sets. 
Working in the continuum space is a possibility, but the calculation would be much more involved than in molecular systems.
Second is the systematic improvement of the calculations for excited states.
\textcolor{black}{It is intriguing to investigate the quantitative performance; whether higher-order subspace expansions can be efficiently implemented, how the accuracy is compared to other excited-state calculation framework such as the equation-of-motion and time-dependent linear response~\cite{mussard_2018}, and so on.}
Third is the behaviour of physical observables.
One may want to know the optical/magnetoelectric/thermal responses, so that experimental results can be directly compared.
If the system is either quasi-static or static, those properties can be evaluated as derivatives of the energy with respect to an external perturbation (e.g., electric field)~\cite{Pulay2014}.

The main bottleneck that prevents the simulation by the NQS in larger systems is the sampling efficiency.
As mentioned by Choo {\it et al.} for the case of RBM~\cite{choo_2020}, and as known before in the VMC community, accurate calculations for relatively weak electronic correlations in the HF basis requires increasingly larger number of Monte Carlo samplings, because the amplitudes for multi-electron excitations are small.
One may consider applying efficient sampling techniques, such as parallel tempering, heat-bath configuration interaction~\cite{holmes_2016}, or even employ non-HF bases.
\section{Data Availability}
The data that support the findings of this study are available
from the corresponding author upon request.

\section{Code Availability}
Codes written for and used in this study is available from the corresponding author upon reasonable request.

\section{Acknowledgements}
We thank Kenny Choo, Antonio Mezzacappo, and James Spencer for fruitful discussions. 
This work was supported by MEXT Quantum Leap Flagship Program (MEXT Q-LEAP) Grant Number JPMXS0118067394 and JPMXS0120319794.
N.Y. is supported by the Japan Science and Technology Agency (JST) (via the Q-LEAP program).
W.M. wishes to thank Japan Society for the Promotion of Science (JSPS) KAKENHI No.\ 18K14181 and JST PRESTO No.\ JPMJPR191A.
F.N. is supported in part by: NTT Research,
Army Research Office (ARO) (Grant No. W911NF-18-1-0358),
Japan Science and Technology Agency (JST)
(via the CREST Grant No. JPMJCR1676),
Japan Society for the Promotion of Science (JSPS) (via the KAKENHI Grant No. JP20H00134
and the JSPS-RFBR Grant No. JPJSBP120194828),
the Asian Office of Aerospace Research and Development (AOARD) (via Grant No. FA2386-20-1-4069),
and the Foundational Questions Institute Fund (FQXi) via Grant No. FQXi-IAF19-06.
Numerical calculations were performed using OpenFermion~\cite{openfermion}, PySCF (v1.7.1)~\cite{pyscf}, and  NetKet~\cite{netket}.
Some calculations were performed using the supercomputer systems in RIKEN (HOKUSAI GreatWave), the Institute of Solid State Physics at the University of Tokyo, 
and in the Research Institute for Information Technology (RIIT) at Kyushu University, Japan.

\section{Author contributions}
N.Y. and W.M. conceived the project and contributed equally to the numerical simulations.
W.M. and F.N. supervised the research.
All authors discussed the results and contributed to writing the paper.

\section{Competing interests}
The authors declare no competing interests.

\section{Methods}
\subsection{Stochastic imaginary-time evolution by variational Monte Carlo}\label{app:imaginary_time}
Given an initial state $\ket{\Psi_0}$ whose overlap with the true ground state is nonzero (and desirably not exponentially small), the ground state $\ket{\Psi_{\rm GS}}$ can be simulated as 
\begin{eqnarray}\label{eqn:imag_time_app}
\ket{\Psi_{GS}} \propto \lim_{N \rightarrow \infty}\lim_{\eta \rightarrow 0}
    \left(\prod_{k=1}^N e^{-\eta H}\right)\ket{\Psi_0},
\end{eqnarray}
where $H$ is the Hamiltonian of the system and $\eta$ is a "learning rate" that determines the step of the imaginary-time evolution.
The exact simulation of Eq.~\eqref{eqn:imag_time_app} for generic quantum many-body systems becomes exponentially inefficient as the system size grows. 
Hence, we approximate the quantum state by a variational ansatz $\ket{\Psi_{\theta}}$ and consider the update rule of the parameters $\theta$ such that Eq.~\eqref{eqn:imag_time_app} is realized approximately.

There are numerous variational principles that dictate the parameter updates. 
Here, we choose the stochastic reconfiguration method~\cite{amari_1992, sorella_2001}, which uses the Fubini-Study metric $\mathcal{F}$ to measure the difference between the exact and variational imaginary-time evolution. 
Given a set of variational parameter $\theta$, the update $\delta \theta$ is determined as 
\begin{eqnarray}
    \delta \theta &=& \mathop{\rm arg~min}\limits_{\Delta}
    \left(\mathcal{F}\left[
    e^{-\eta \hat{H}} \ket{\Psi_{\theta}}, \ket{\Psi_{\theta + \Delta}}
    \right]
    \right) \nonumber\\
    &=& -\eta g^{-1} f \label{eqn:sr}
\end{eqnarray}
where $\mathcal{F}[\ket{\psi}, \ket{\phi}] = \arccos(\sqrt{\braket{\psi|\phi} \braket{\phi|\psi}/\braket{\psi|\psi} \braket{\phi|\phi}})$ and
elements of the generic force $f_i$ and the geometric tensor $g_{ij}$ are given as 
\begin{eqnarray}
    f_i &=&  \frac{\braket{\partial_i\Psi_{\theta}|H|\Psi_{\theta}}}{\braket{\Psi_{\theta}|\Psi_{\theta}}} - \frac{\braket{\partial_i\Psi_{\theta}|\Psi_{\theta}}}{\braket{\Psi_{\theta}|\Psi_{\theta}}} \frac{\braket{\Psi_{\theta}|H|\Psi_{\theta}}}{\braket{\Psi_{\theta}|\Psi_{\theta}}}, \\
    g_{ij} &=& \frac{\braket{\partial_i \Psi_{\theta}|\partial_j \Psi_{\theta}}}{\braket{\Psi_{\theta}|\Psi_{\theta}}} - \frac{\braket{\partial_i \Psi_{\theta}|\Psi_{\theta}}}{\braket{\Psi_{\theta}|\Psi_{\theta}}} \frac{\braket{\Psi_{\theta} | \partial_j \Psi_{\theta}}}{\braket{\Psi_{\theta}|\Psi_{\theta}}},
\end{eqnarray}
where $\partial_i$ is the derivative with respect to the $i$-th element of the parameter $\theta_i$.
\textcolor{black}{
It is noteworthy that the geometric tensor $g$ is the extension of the Fisher information to quantum states. 
The stochastic gradient method based on $g$, or the Fisher information, was independently developed in the machine learning community~\cite{amari_1992}, and is frequently referred  to as the natural gradient method.
}

\textcolor{black}{Note that both $f$ and $g$ can be estimated efficiently using Monte Carlo sampling.
Indeed, any physical observable $O$ can be estimated for a quantum state $\ket{\Psi}$ as
\begin{eqnarray}
    \braket{O} = \frac{\braket{\Psi|O|\Psi}}{\braket{\Psi|\Psi}}
    = \frac{\sum_{\sigma} |\Psi(\sigma)|^2 O_{\rm loc}(\sigma)}{\sum_{\sigma} |\Psi(\sigma)|^2}
     = \sum_{\sigma} p(\sigma) O_{\rm loc}(\sigma),\nonumber \\
     \  
\end{eqnarray}
where $O_{\rm loc}(\sigma) = \sum_{\sigma'}\frac{\Psi(\sigma')}{\Psi(\sigma)} \braket{\sigma|O|\sigma'}$ is introduced to enable the simulation of the expectation value from classical sampling over the probability distribution $p(\sigma) = |\Psi(\sigma)|^2/\sum_{\sigma}|\Psi(\sigma)|^2$.
Using the Metropolis--Hastings algorithm with particle number conservation, we typically sample $\bigO(10^{5})$ to $\bigO(10^7)$ spin configurations to estimate $p(\sigma)$. 
Each configuration is drawn every 10 to 20 Monte Carlo steps so that the autocorrelation, and hence the sampling error, is sufficiently small when the optimization converges.
}

Three technical remarks are in order.
\textcolor{black}{
First, we take the initial state $\ket{\Psi_0} (=\ket{\Psi_{\theta_0}^{\rm RBM}})$ as the HF state such that the overlap with the ground state is non-zero. 
Small noise is added to avoid the gradient vanishing problem, which arises when the parameters of the RBM state are tuned to express any computational basis exactly.
}
Second, to stabilize the optimization, small number $\epsilon$ is uniformly added to the diagonal elements of $g$ as $g_{ii} \rightarrow g_{ii} + \epsilon$. 
While large $\epsilon$ is beneficial in early iterations, it is necessary to decrease it, or otherwise one may result in undesirable local minima.
Therefore, $\epsilon$ is initially set as $\mathcal{O}(10^{-2})$ and gradually decreased to $\mathcal{O}(10^{-3})$ after several hundred steps. 
Third, we find that it is crucial to adopt an appropriate scheduling of $\eta$ to speed up the optimization and, more importantly, avoid local minima. 
In the present work, we exclusively employ the RMSProp method~\cite{hinton_rmsprop_2012}, which adaptively modifies $\eta$ according to the magnitude of the gradient.

\subsection{Energy corrections by the extended Koopmans' theorem}\label{app:correction}
\begin{figure}[t]
\begin{center}
    \includegraphics[width=0.95\columnwidth]{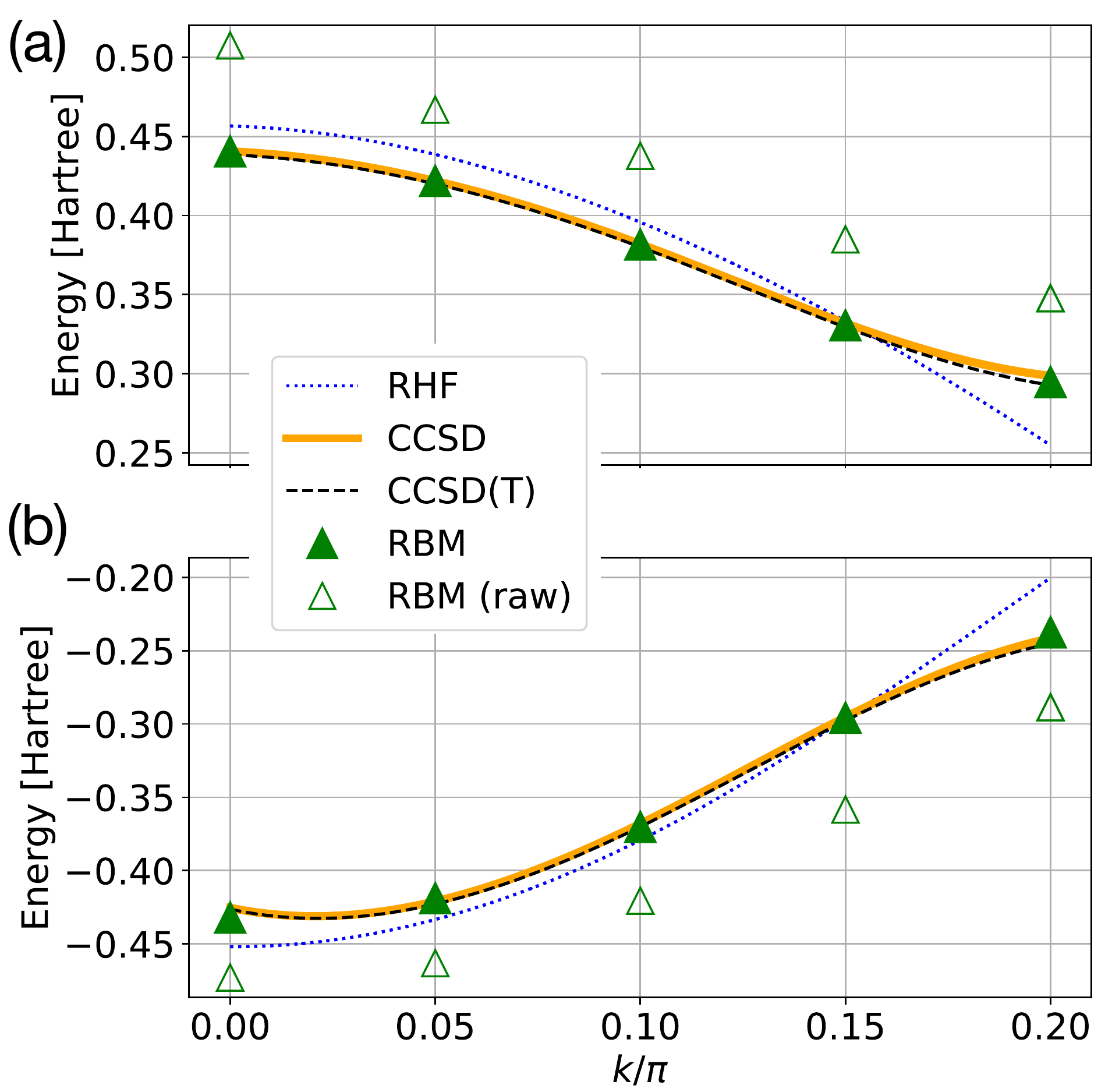}
\caption{\label{fig:C2H2_rawband} {\bf The effect of energy correction to quasiparticle bands.} Here, we display the (a) lowest conduction band and (b) highest valence band.  The unfilled green triangle denotes the raw values obtained by solving Eq.~\eqref{eqn:qse} defined in the main text, and the filled ones indicate the values corrected by Eq.~\eqref{eqn:koopmans} following the extended Koopmans' theorem.
The blue dotted lines, yellow lines, and black dashd lines indicate the result by restricted Hartree-Fock method, coupled-cluster equation-of-motion formalism with singles and doubles (EOM-CCSD), and EOM-CCSD with perturbative triple excitation (EOM-CCSD(T)(a)*), respectively.
}
\end{center}
\end{figure}

\textcolor{black}{
In Fig.~\ref{fig:C2H2_rawband}, we visualize the effect of the corrections to the energy bands by the extended Koopmans' theorem, which are defined in Eq.~\eqref{eqn:koopmans} in the main text as
\begin{eqnarray}
\begin{cases}
\Delta E^{IP} &= E_{GS}^{N_v} - E_{GS}^{N_v-1}, \nonumber\\
\Delta E^{EA} &= E_{GS}^{N_v+1} - E_{GS}^{N_v},\nonumber
\end{cases}
\end{eqnarray}
where $E_{GS}^{n}$ is the energy of the RBM optimized in the particle-number sector $n$. Here, panels (a) and (b) indicate the first conduction and valence bands, respectively.
In both bands, we observe a systematic deviation, which we attribute to the lack of orbital relaxation effect caused by the removal or addition of a single electron.
The order of the correction $\Delta E \sim 0.05$ Ha is comparable to that of the electronic correlation ($\sim 0.1$ Ha).
}

\section{References}
\bibliography{bib_yoshioka}
\end{document}